\documentclass[12pt]{article}
\pdfoutput=1
\usepackage[utf8]{inputenc}
\usepackage{amsmath}
\usepackage{amssymb}
\usepackage{units}
\usepackage{graphicx}
\usepackage{slashed}
\usepackage{array}
\usepackage[leftcaption,raggedright]{sidecap}
\usepackage{caption}
\usepackage[hidelinks]{hyperref}

\newcommand\pubnumber{NuPhys2026-Nick-Latham}
\newcommand\pubdate{\today}

\def\napoli{on behalf of the T2K collaboration}

\def\support{\footnote{
  King's College London.
}}

\def\Title#1{\begin{center} {\Large #1 } \end{center}}
\def\Author#1{\begin{center}{ \sc #1} \end{center}}
\def\Address#1{\begin{center}{ \it #1} \end{center}}

\newcommand\pubblock{\rightline{\begin{tabular}{l} \pubnumber\\
         \pubdate  \end{tabular}}}
\newenvironment{Abstract}{\begin{quotation}  }{\end{quotation}}
\newenvironment{Presented}{\begin{quotation} \begin{center} 
             PRESENTED AT\end{center}\bigskip 
      \begin{center}\begin{large}}{\end{large}\end{center} \end{quotation}}

\def\beq{\begin{equation}}
\def\eeq#1{\label{#1}\end{equation}}
\def\eeqn{\end{equation}}

\def\beqa{\begin{eqnarray}}
\def\eeqa#1{\label{#1}\end{eqnarray}}
\def\eeqan{\end{eqnarray}}

\let\bar=\overbar

\def\Dslash{\not{\hbox{\kern-4pt $D$}}}
\def\dslash{\not{\hbox{\kern-2pt $\del$}}}

\def\msb{{\bar{\ssstyle M \kern -1pt S}}}

\begin{document}
\begin{titlepage}
\pubblock

\vfill
\Title{Recent Neutrino Oscillation and Cross-Section Results from the T2K Experiment}
\vfill
\Author{ Nick Latham\support}
\Address{\napoli}
\vfill
\begin{Abstract}
The T2K long-baseline neutrino oscillation experiment in Japan continues to lead the search for leptonic charge-parity violation while providing precision measurements of mixing and mass splitting parameters. Central to this programme is the mitigation of systematic uncertainties through the near detector complex, which provides high-statistics neutrino–nucleus interaction cross-section measurements across various targets. This contribution presents the latest T2K oscillation results, incorporating the first data with a gadolinium-loaded far detector, and highlights several recent cross-section measurements, including several world-first measurements of rare interaction channels. Together, these results demonstrate the vital synergy between interaction modelling and oscillation analysis in the search for charge-parity violation in the T2K-II era.
\end{Abstract}
\vfill
\begin{Presented}
NuPhys2026, Prospects in Neutrino Physics\\
King's College, London, UK,\\ January 7--9, 2026
\end{Presented}
\vfill
\end{titlepage}
\def\thefootnote{\fnsymbol{footnote}}
\setcounter{footnote}{0}

\section{Introduction}

The T2K experiment~\cite{T2K} provides world-leading constraints on neutrino oscillation parameters using a high-intensity off-axis muon (anti)neutrino beam. The physics programme relies on two primary channels: the disappearance of $\nu_\mu$ ($\bar{\nu}_\mu$), which yields precise measurements of the atmospheric parameters $\sin^2\theta_{23}$ and $|\Delta m^2_{32}|$, and the appearance of $\nu_e$ ($\bar{\nu}_e$). The latter is sensitive to the mixing angle $\theta_{13}$ and enables measurements of the charge-parity ($CP$)-violating phase, $\delta_{\mathrm{CP}}$, through the comparison of neutrino and antineutrino oscillation probabilities. These measurements also provide sensitivity to the neutrino mass ordering through matter effects and the interplay between oscillation parameters.

Reaching the sensitivity threshold for high-precision oscillation parameter determination demands a rigorous treatment of systematic uncertainties, which are currently dominated by neutrino--nucleus interaction modelling. Because T2K utilises the kinematics of the final-state lepton to reconstruct the neutrino energy, nuclear effects such as multi-nucleon correlations and final-state interactions can bias the inferred oscillation parameters. The T2K near detector complex is therefore utilised to perform high-statistics cross-section measurements, as well as to study the initial beam composition. These measurements are essential to constrain the underlying interaction models and characterise the neutrino flux, providing the necessary foundation for precise oscillation analysis.

\section{The T2K Experiment}

\subsection{Beamline and Far Detector}

The initial $\nu_\mu$ beam is produced at the J-PARC accelerator facility in Tōkai, Japan. The beam is directed $295$~km west towards Super-Kamiokande (SK)~\cite{Super-K} in Kamioka, a 50~kt water Cherenkov tank which serves as the far detector of the T2K experiment. Neutrino and antineutrino beam modes are selected by reversing the polarity of a set of three magnetic horns using forward or reverse horn current respectively. The neutrino beam is focused such that it propagates $2.5^{\circ}$ off-axis with respect to SK, producing a narrow-band neutrino energy spectrum peaked at approximately $0.6$~GeV, corresponding to the first oscillation maximum for the T2K baseline.

The primary proton beam at J-PARC has recently undergone significant upgrades, achieving a stable power of $750$~kW~\cite{NDUP} by increasing the repetition rate and upgrading the main ring power supplies. The three magnetic horns now operate at a current of $320$~kA to further enhance the flux.

Since 2022, the ultra-pure water in SK has been loaded with $0.03\%$ gadolinium sulfate octahydrate by mass~\cite{Gd}. This upgrade enables high-efficiency thermal neutron tagging via the radiative capture on Gd, which yields an ${\sim}8$~MeV $\gamma$-ray cascade. Neutron tagging serves as a new tool for discriminating between neutrino and antineutrino interactions and suppressing atmospheric neutrino backgrounds.

\subsection{Near Detector Complex}

The T2K near detector complex, located $280$~m downstream of the production target, characterises the unoscillated neutrino beam and constrains interaction models. The suite primarily consists of the off-axis ND280 and WAGASCI--BabyMIND detectors, which provide the high-statistics data necessary to reduce systematic uncertainties in the far detector oscillation analysis.

The ND280 detector is a magnetised tracking spectrometer. At its core, two Fine-Grained Detectors (FGDs)~\cite{FGD} serve as the primary interaction targets (carbon/hydrogen and water, respectively). These are interleaved with three Time Projection Chambers (TPCs)~\cite{TPC} that provide precise tracking and particle identification via ionisation energy loss. The entire sub-detector system is housed within the UA1/NOMAD magnet, providing a $0.2$~T magnetic field for charge and momentum determination.

Complementary measurements are provided by the WAGASCI--BabyMIND detectors~\cite{WAGASCI}. WAGASCI utilises a 3D grid of plastic scintillator bars surrounding water targets to achieve nearly $4\pi$ angular acceptance. The downstream BabyMIND magnetised spectrometer facilitates muon charge and momentum reconstruction. This configuration is optimised for water cross-section measurements and more closely matches the angular phase space of SK.

\subsection{ND280 Upgrade}

T2K has recently completed a major upgrade of the ND280 detector. While the oscillation and cross-section results presented in this contribution utilise the baseline configuration, the upgraded detector is now taking data and will provide improved constraints for future analyses. The upgrade replaces the upstream portion of the detector with a high-granularity scintillator target (SuperFGD), two High-Angle TPCs (HA-TPCs), and a suite of Time-of-Flight (TOF) detectors. This new configuration is designed to significantly improve the detection efficiency for high-angle and backward-going tracks, while lowering the threshold for proton and pion reconstruction.

\begin{figure}[htbp]
\centering
\includegraphics[width=0.99\textwidth]{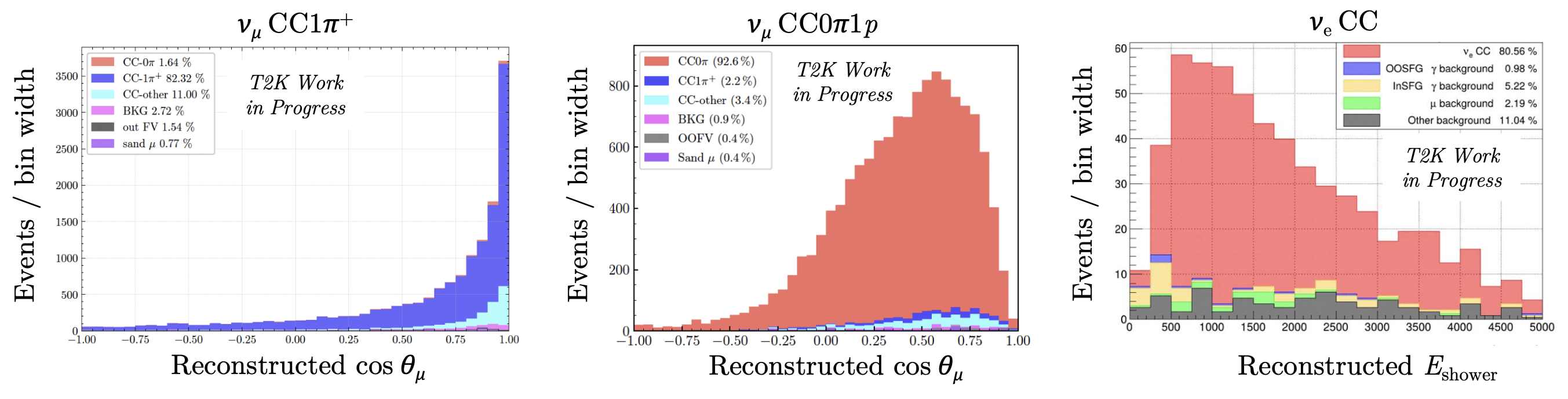}
\caption{Example event selection distributions obtained using the upgraded ND280 detector configuration. The plots show reconstructed kinematic variables for three representative selections: $\nu_\mu\mathrm{CC}1\pi$ (left), $\nu_\mu\mathrm{CC}0\pi 1p$ (centre), and $\nu_e\mathrm{CC}$ (right). The stacked histograms indicate the predicted signal and background composition for each sample. These plots are adapted from Ref.~\cite{NDUP}.}
\label{fig:NDUP_results}
\end{figure}

As of early 2026, the first data from the upgraded ND280 detector is being analysed to optimise selection efficiencies and evaluate detector systematic uncertainties. Preliminary studies (Fig.~\ref{fig:NDUP_results}) indicate substantial improvements across near detector samples; notably, the photon background in the $\nu_e\mathrm{CC}$ selection is greatly reduced from ${\sim}30\%$ in the previous inclusive analysis~\cite{nueCC}. Furthermore, the $\nu_\mu\mathrm{CC}0\pi$ and $\nu_\mu\mathrm{CC}1\pi^+$ samples are able to reach purities of approximately $90\%$, including for backwards-going tracks $\left( \cos{\theta_\mu} < 0 \right)$ due to the improved angular acceptance of ND280 post upgrade. These improvements are expected to enhance the constraining power of ND280 measurements in future oscillation analyses.

\section{Latest Oscillation Results}

\subsection{Standalone T2K Results}

The latest T2K oscillation analysis incorporates an updated dataset totalling $21.4 \times 10^{21}$ protons on target (POT); this is a 9\% increase in $\nu$-mode data compared to the previous analysis~\cite{OA_previous}. A significant milestone in this analysis is the inclusion of the first results from the gadolinium-loaded SK detector. The enhanced neutron-tagging efficiency provided by the Gd loading has allowed for refined far-detector event selections for Michel electrons. In particular, the $\nu_e\mathrm{CC}1\pi^{+}$ sample has undergone a renewed assessment of systematic uncertainties, including a new treatment of the pion-nucleus interactions to improve the overall model robustness.

\begin{figure}[htbp]
\centering
\includegraphics[width=0.99\textwidth]{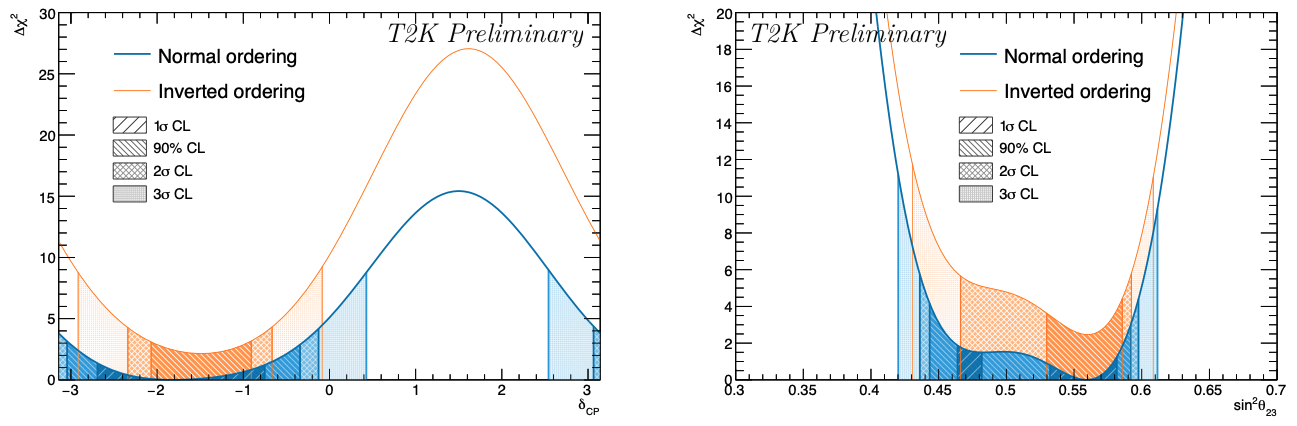}
\caption{Frequentist likelihood scans of the oscillation parameters $\delta_{\mathrm{CP}}$ (left) and $\sin^2\theta_{23}$ (right) from the latest T2K oscillation analysis. The curves show the $\Delta\chi^2$ profiles for the normal ordering (blue) and inverted ordering (orange) hypotheses. The shaded regions indicate the corresponding confidence intervals at the $1\sigma$, $90\%$, $2\sigma$, and $3\sigma$ levels. These plots are taken from Ref.~\cite{Neutrino2024_T2K}.}
\label{fig:OA2023_results}
\end{figure}

In the latest results~\cite{Neutrino2024_T2K}, $CP$ conservation is excluded at the 90\% confidence level in the nominal analysis. Robustness-of-fit studies, involving 18 alternative interaction and systematic model variations, demonstrate that this exclusion remains stable across the majority of tested scenarios. The data shows a subtle preference for the Normal Ordering (NO) over Inverted Ordering (IO), with a Bayes factor of $\mathrm{NO/IO} = 3.3$. Similarly, a slight preference for the upper $\theta_{23}$ octant is observed, with Bayes factor 2.6. The precision on the atmospheric mass splitting reaches 2\% at the 1$\sigma$ level with central value $|\Delta m^2_{32}| \approx 2.5 \times 10^{-3}$~eV$^2$, assuming the NO. The frequentist likelihood scans of $\delta_{\mathrm{CP}}$ and $\sin^{2}{\theta_{23}}$ are shown in Fig.~\ref{fig:OA2023_results}.

\subsection{Joint Fits}

Beyond the T2K-only results, two joint analyses with external datasets provide additional insight into the current preferred oscillation parameter values. These joint fits are essential for breaking parameter degeneracies, such as the entanglement of the mass ordering and $\delta_{\mathrm{CP}}$ effects, by combining data from different baselines and neutrino energy spectra. One joint fit combines T2K data with input from the NOvA~\cite{NOvA} experiment, enhancing sensitivity through its significantly longer $810$~km baseline and higher beam energy of $2$~GeV, which amplifies the impact of Earth-matter effects on measuring oscillation probabilities. The second joint fit uses SK atmospheric data~\cite{T2KSKJoint} which provides a unique probe of the mass ordering via the high-statistics sample of upward-going neutrinos that traverse the Earth's inner structure.

The T2K--SK joint fit~\cite{T2KSKJoint} finds a subtle preference for the NO and excludes $CP$-conserving values of the Jarlskog invariant at the $1.9\text{--}2.0\sigma$ level. Conversely, the T2K--NOvA joint analysis~\cite{T2KNOvAJoint} exhibits a subtle preference for the IO. This combined fit achieves $1.4\%$ precision on $|\Delta m^2_{32}|$ and finds that $CP$ conservation is excluded at the $3\sigma$ level for the inverted ordering, while a broader range of $\delta_{\mathrm{CP}}$ values remains permissible under the normal ordering. In both joint analyses, the $\theta_{23}$ octant ambiguity persists.

\section{Recent Cross-Section Measurements}

The T2K near detectors enable a broad programme of neutrino–nucleus cross-section measurements across a range of interaction channels and nuclear targets. These measurements probe interaction processes relevant to oscillation analyses and provide constraints on neutrino interaction models at the T2K beam energy. Cross sections are reported in terms of final-state particle topologies rather than underlying interaction modes, since nuclear effects can obscure the true underlying process. This section summarises the most recent cross-section measurements performed using the near detectors.

\subsection{$\boldsymbol{\nu_e} \mathrm{\textbf{CC}} \boldsymbol{\pi^+}$ \textbf{on Carbon}}

Improved measurements of $\nu_e$ interaction channels are particularly important for oscillation analyses, where modelling uncertainties can impact the reconstruction of the neutrino energy spectrum. Electron-neutrino charged-current pion production $\left( \nu_e \mathrm{CC}\pi^+ \right)$ is a crucial channel for studies of $\nu_e$ appearance at the T2K far detector. T2K recently performed the first measurement of this process on a carbon target $\left( \nu_e A \rightarrow e^- \pi^+ X \right)$ using ND280~\cite{nueCCPi}. Differential cross sections are measured as a function of the outgoing electron and pion kinematics $\left( p_e, \theta_e, p_\pi \right)$. 

In addition to standard track reconstruction using the TPCs, this analysis employs a novel reconstruction technique to reconstruct low momentum pions which are below the ND280 tracking threshold. Such pions typically decay within the FGDs and the hit patterns associated with the Michel electron from the decay chain $\pi^+ \rightarrow \mu^+ \rightarrow e^+$ are used to approximate pion momentum. This approach substantially improves sensitivity to the cross section in the $p_\pi < 200$~MeV$/c$ region, which has previously been difficult to access in ND280 pion production measurements.

\begin{figure}[htbp]
\centering
\includegraphics[width=0.99\textwidth]{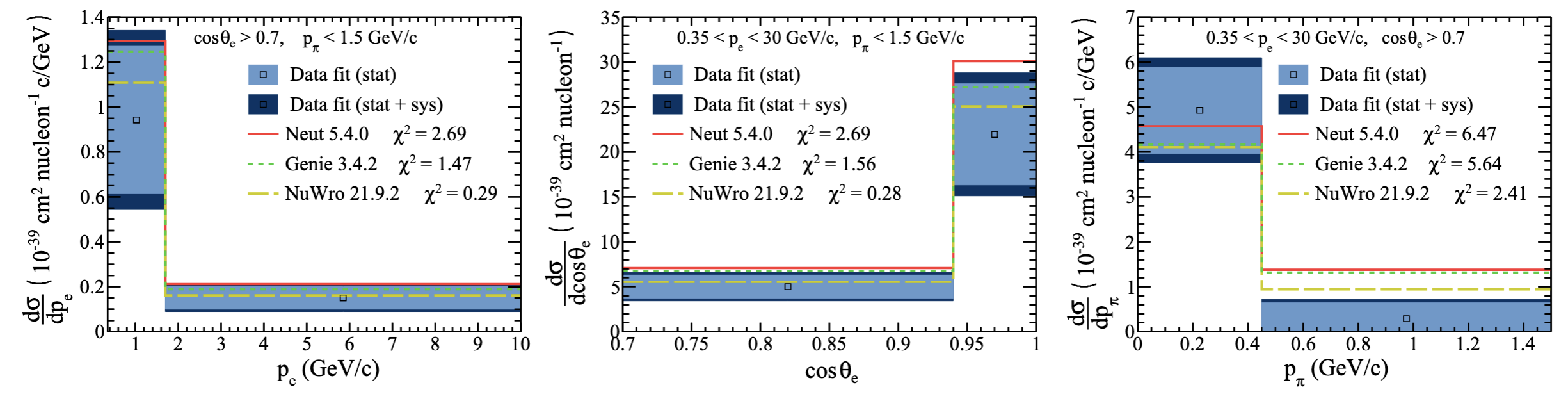}
\caption{Differential cross-section measurements and event generator predictions as a function of final-state particle kinematics for $\nu_e \text{CC} \pi^+$ on carbon. These plots are taken from Ref.~\cite{nueCCPi}.}
\label{fig:nueCCPi_results}
\end{figure}

Comparisons of the measured differential cross sections with neutrino interaction generators (Fig.~\ref{fig:nueCCPi_results}) show tensions in certain regions of phase space. In particular, predictions from Neut~5.4~\cite{NEUT} and Genie~3.4~\cite{GENIE} exhibit discrepancies at the level of $2\text{--}3\sigma$ for the highest-momentum pions in the sample $\left( p_\pi > 1.5~\text{GeV}/c \right)$, suggesting that current resonant or non-resonant pion production models, or the treatment of the transition to the deep inelastic scattering regime, may require further refinement.

\subsection{$\mathrm{\textbf{NC}} \boldsymbol{1\pi^+}$ \textbf{on Carbon}}

Neutral-current single charged pion production $\left(\mathrm{NC}1\pi^+\right)$ constitutes an important background to the muon-neutrino disappearance sample at the T2K far detector. In these interactions $\left( \nu N \rightarrow \nu \pi^+ N' \right)$ the outgoing neutrino is undetected and the visible final state is typically dominated by a single positively charged pion, making this channel particularly sensitive to final-state interactions and nuclear effects. In a water Cherenkov detector such as SK, the Cherenkov ring produced by a charged pion can in some cases be misidentified as a muon-like ring. Precise measurements of $\mathrm{NC}1\pi^+$ production are therefore important for constraining this background in oscillation analyses.

\begin{figure}[htbp]
\centering
\includegraphics[width=0.99\textwidth]{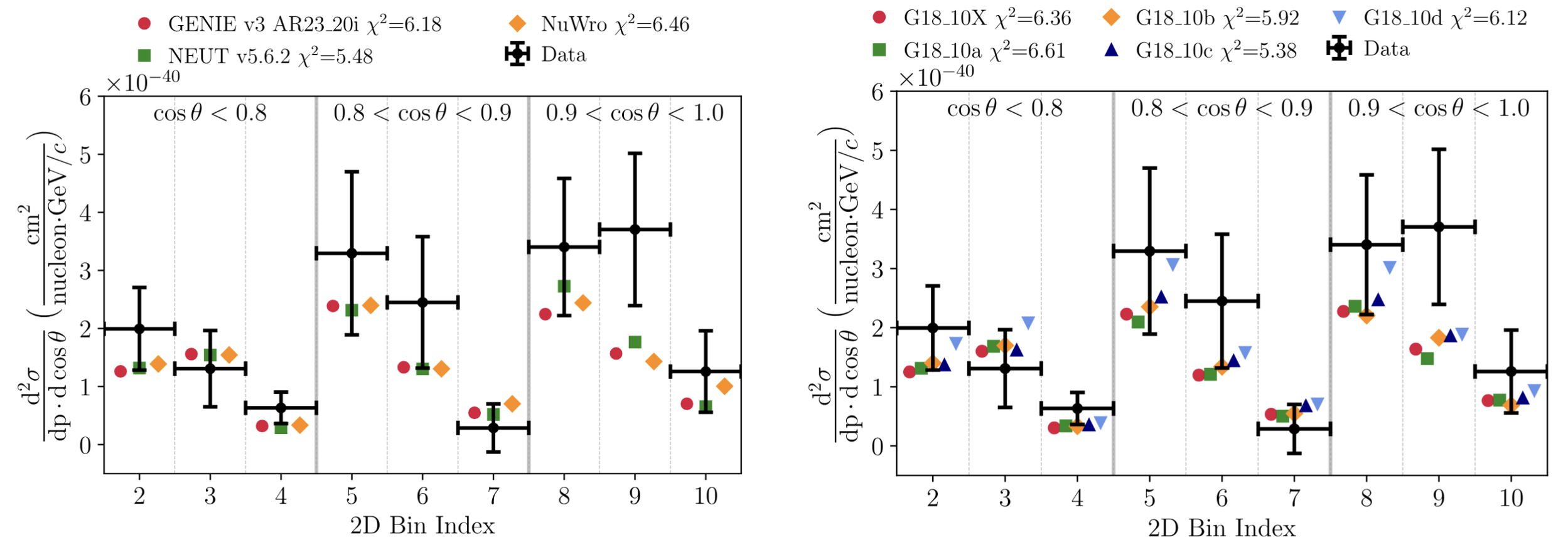}
\caption{Differential cross-section measurements and event generator predictions as a function of final-state pion kinematics for $\text{NC} 1\pi^+$ on carbon. The left plots show comparisons with standard event generators, the right plot shows comparisons with various specific final-state interaction models. These plots are taken from Ref.~\cite{NC1Pi}.}
\label{fig:NC1Pi_results}
\end{figure}

T2K has recently performed a measurement of this channel on carbon using the ND280 near detector, reporting differential cross sections as functions of the pion kinematics~\cite{NC1Pi}. The differential cross sections (Fig.~\ref{fig:NC1Pi_results}) are generally under-predicted by simulations, with an average discrepancy of ${\sim}30\%$. 

\subsection{New $\boldsymbol{\nu_\mu} \mathrm{\textbf{CC}} \boldsymbol{0\pi}$ Measurements}

Charged-current pion-less interactions involving muon neutrinos $\left(\nu_\mu\mathrm{CC}0\pi\right)$ constitute the dominant interaction channel at the T2K beam energy. Precise constraints on this process are important for oscillation analyses, as $\nu_\mu\mathrm{CC}0\pi$ events form the primary signal sample for muon-neutrino disappearance at the far detector.

\begin{figure}[htbp]
\centering
\includegraphics[width=0.99\textwidth]{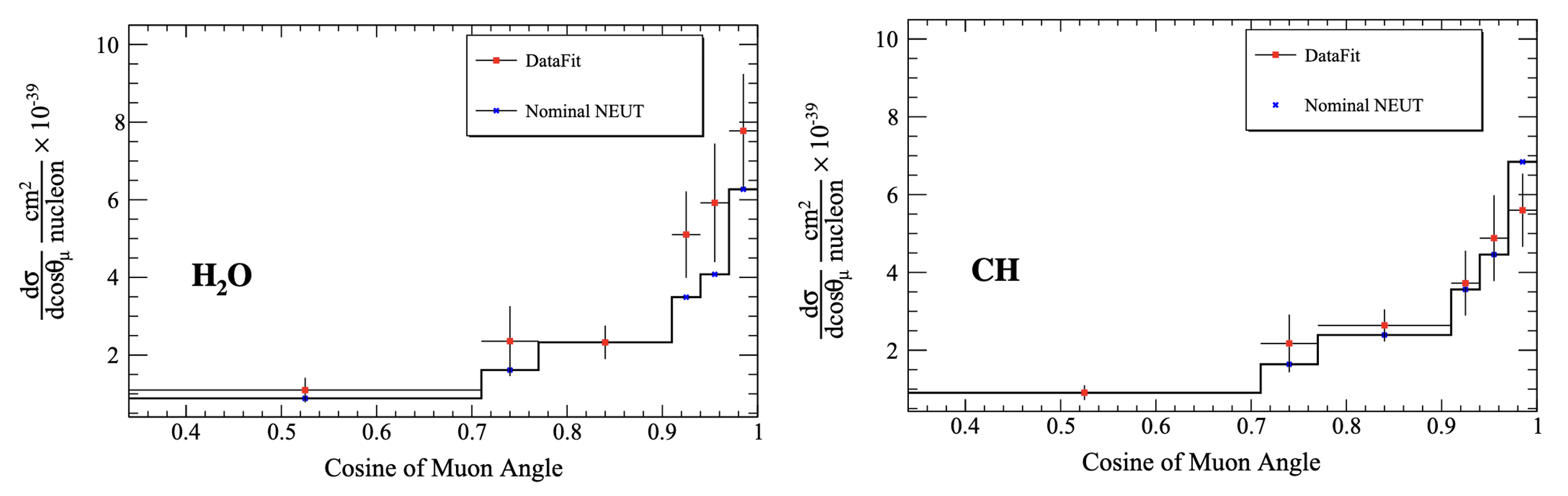}
\caption{Differential cross-section measurements and event generator predictions as a function of final-state muon angle with respect to the neutrino beam for $\nu_\mu \text{CC} 0\pi$ on water and carbon. These plots are taken from Ref.~\cite{WAGASCI_numuCC0Pi}.}
\label{fig:WAGASCI_results}
\end{figure}

T2K has performed a measurement of $\nu_\mu\mathrm{CC}0\pi$ interactions on both water and carbon targets using the WAGASCI--BabyMIND detectors~\cite{WAGASCI_numuCC0Pi}. The WAGASCI detector provides a finely segmented target with nearly isotropic acceptance for charged particles, while the downstream BabyMIND spectrometer measures the momentum and charge of outgoing muons. This measurement represents the first determination of the $\nu_\mu\mathrm{CC}0\pi$ cross section on water at the T2K near detector site with full angular coverage. Differential cross sections as a function of muon kinematics are measured and compared with simulations for both water and hydrocarbon targets (Fig.~\ref{fig:WAGASCI_results}); good agreement is seen in most bins. The results provide an important validation of interaction modelling on water, the target material of SK.

In addition, T2K has recently performed a measurement of $\nu_\mu\mathrm{CC}0\pi$ interactions with one or more protons in the final state on carbon and oxygen targets~\cite{T2K_CC0pi_TKI}. Differential cross sections are reported using transverse kinematic imbalance (TKI) variables~\cite{TKI, Lu, Furmanski}; these decompose the transverse momentum imbalance into magnitude and angular components to explicitly isolate Fermi motion from final-state interactions and multi-nucleon effects. The use of TKI variables allows nuclear effects to be directly probed while minimising neutrino energy reconstruction bias. Differential cross-section measurements as a function of TKI variables indicate that a range of neutrino event generators, with different final-state interaction configurations, struggle to consistently describe the interaction across all regions of phase space.

\section{Summary and Outlook}

T2K continues to play a leading role in the search for leptonic $CP$ violation, with $CP$ conservation currently excluded at the 90\% confidence level in the nominal analysis. This result remains robust under a wide range of alternative interaction and systematic model variations. In parallel, the near detector programme has produced several world-first differential cross-section measurements of sub-dominant neutrino interaction channels. Three of the four most recent cross-section measurements show tensions with current interaction model predictions, suggesting that further model refinement may be required, although the statistical uncertainties remain large.

Future developments include the development of existing near detector samples using the upgraded ND280, the implementation of new near detector samples featuring tagged neutrons, $\bar{\nu}_\mu\mathrm{CC}\gamma$ interactions, and selections with full $4\pi$ solid-angle acceptance. Work is also underway to formally include WAGASCI--BabyMIND data in the oscillation fits. At the far detector, new samples are under development, including enhanced $\nu_e\mathrm{CC}1\pi^{+}$ and $\mathrm{NC}1\pi^0$ selections that leverage the improved neutron-tagging capabilities of SK-Gd. Complementing these detector-side improvements, the J-PARC beamline continues to scale toward a target power of $1.3$~MW. Together, these advancements ensure that T2K remains at the forefront of precision measurements for neutrino oscillations.

\bibliographystyle{unsrt}
\bibliography{references}

\end{document}